# Tuple-based abstract data types: full parallelism

Carmelo MALTA, José MARTINEZ

Université des Sciences et Techniques du Languedoc
Laboratoire de Systèmes Informatiques
860, rue de Saint-Priest
34090 Montpellier, FRANCE
E-mail : <malta, martinez>@crim.fr

**Abstract**
Commutativity has the same inherent limitations as compatibility. Then, it is worth conceiving simple concurrency control techniques. We propose a restricted form of commutativity which increases parallelism without incurring a higher overhead than compatibility. Advantages of our proposition are: (1) commutativity of operations is determined at compile-time, (2) run-time checking is as efficient as for compatibility, (3) neither commutativity relations, (4) nor inverse operations, need to be specified, and (5) log space utilization is reduced.

## 1. INTRODUCTION

Shared data are manipulated through *transactions* [4]. A transaction is a unit of consistency, i. e., from a consistent state of the database, it delivers another consistent state. The interleaved execution of a set of transactions is controlled in order to maintain this consistency property. The main (syntactic) criterion is *serializability*: an interleaved execution must be equivalent to any serial execution.

Accesses to individual items obey a local property which is generally *compatibility*, i. e., operations are either writers, executed in mutual exclusion, or readers, executed in parallel exclusively with other readers. To enhance performances, several authors exploit *commutativity* of operations on abstract data types (ADTs), a *generalization* of compatibility [15, 17].

Commutativity classifies operations in more categories than just readers and writers. Thus, more parallel executions are admissible. However, three points must be enlightened: (1) these studies focus on few data structures: directories [15], maps [16], sets [17], counters [14]; (2)

This work was supported in part by the PRCs BD3 and $C^3$ coordinated by the Centre National de la Recherche Scientifique (CNRS), and in part by the Ministère de la Recherche et de la Technologie (MRT).

commutativity relations are explicitly given; (3) recovery may require inverse operations. One can criticize each of these points: (1) a programmer creates various data structures, with unpredictable, possibly changing, and often simple commutativity properties; (2) it is unthinkable to put the burden of determining the commutativity of each pair of operations on the application programmer for each ADT that he or she creates; (3) similarly, it is not always reasonable to ask him or her to provide each inverse operation. In addition, for *arbitrary* objects, commutativity was shown to be a weak enhancement [11]: Write accesses are exclusive with compatibility; exclusive operations are not eliminated by commutativity. Read accesses prohibit concurrent updates; commutativity does not allow to update the predicate of the set which contains the initial (and therefore current) value of the object.

We propose a simple technique which eliminates these drawbacks, without totally sacrificing the benefits of commutativity, in the common case of tuple-based ADTs. Nonetheless, it does not preclude the use of sophisticated techniques. Roughly speaking, the technique is as follows: Each *tuple-based* ADT is a cartesian product. To each operation of each ADT, we associate an *access vector* of the same dimension as its cartesian product. Each value composing this vector will denote the most restrictive access mode used by the operation when manipulating the corresponding field. Constructing access vectors is done at compile-time. At run-time, commutativity is controlled by comparing access vectors. To rid *a priori* control of its inherently pessimistic nature, a second control is done, based on dynamic informations; it permits to achieve indirectly *conditional* commutativity. Recovery uses access vectors as projection patterns to take before- and/or after-images of the accessed items.

The organization of the paper is as follows: First, access vectors are defined and strong properties presented. Then, we discuss the use of access vectors for controlling concurrent accesses and recovering from transaction rejects. Next, our approach is compared to similar propositions. In particular, we shall see that access vectors are necessary to obtain, at least, the same parallelism as in relational databases. The conclusion draws attention to future works.

## 2. ACCESS VECTORS

We begin with two basic definitions about ADTs and classical access modes.

*definition 1*
In the sequel, we use to note A an ADT of dimension N: $A = D_1 \times ... \times D_N$, i. e., composed of N (root) fields. The names of the fields in the source code will be noted $field_1$ up to $field_N$.

*definition 2*
We call $c_{MODES}$ the binary compatibility predicate on MODES, given in extension in Table 1, where MODES = {Null,Read,Write} with Null < Read < Write.

Table 1
Classical compatibility relation

|       | Null | Read | Write |
|-------|------|------|-------|
| Null  | yes  | yes  | yes   |
| Read  | yes  | yes  | no    |
| Write | yes  | no   | no    |

The order relation on MODES is directly derived from the compatibility relation [9].

We decided to propose a concurrency control technique based on pure compatibility at the field level. Therefore, at the ADT level, our restricted form of commutativity is a conjunction of compatible accesses to the different fields.

*definition 3*
Let A be an ADT of dimension N, then to each operation OP we associate an *access vector* $DAV_{A,OP} = (m_1, ..., m_N)$ such that:

$m_i$ = Write $\Leftrightarrow$ there exists an assignment "$field_i$ := <expression>" in the code of OP;

$m_i$ = Read $\Leftrightarrow$ there is no such assignment, but "$field_i$" appears in some expression;

$m_i$ = Null $\Leftrightarrow$ "$field_i$" appears nowhere in the code of OP.

*definition 4*
Let OP and OP' be operations on A with the respective access vectors $DAV_{A,OP} = (m_1, ..., m_N)$ and $DAV_{A,OP'} = (m'_1, ..., m'_N)$, then:

$$DAV_{A,OP} \; c \; DAV_{A,OP'} \diagup \forall \, i: \; 1 \leq i \leq N, \; m_i \; c_{MODES} \; m'_i$$

Definition 5 extends commutativity to bags of operations.

*definition 5*
Let $\mathbf{a} = (DAV_{A,OP_i})_{i \in \{1, ..., m\}}$ be a bag of access vectors of operations on A, then $\mathbf{a}$ is *pairwise commutative* if:

$$\forall \, i, j: \; 1 \leq i \leq m, \; 1 \leq j \leq m, \; i \neq j, \; DAV_{A,OP_i} \; c \; DAV_{A,OP_j}$$

From the definitions and Table 1, it follows that in a bag of pairwise commutative operations on a common instance, on each field there is either exactly one writer, or several readers, or neither writers nor readers (See lemma 1.) Since out-parameters are computed from in-parameters, constants, and fields, which cannot be used incompatibly, it is immediate that a bag of pairwise commutative operations can be executed in real parallelism, i. e., without any control! This property justifies the title of this paper.

In other words, c does indeed define a commutativity relation. In point of fact, commutativity of access vectors is a stronger condition than common commutativity [17]: Atomicity of the execution of operations is just one of the conditions in the definition of commutativity, whereas here it is a sufficient condition for operations to be commutative. We call it *strong commutativity*.

### 3. CONTROLLING DIRECT AND INVERSE OPERATIONS

Concurrent accesses are authorized or denied on behalf of commutativity of respective access vectors. Controlling concurrent accesses as efficiently as compatibility is an advantage of this kind of commutativity: concurrent accesses to one instance is controlled in time O(N).

We need two *control vectors* per instance: one to keep track of the number of readers per field, one to keep track of the presence of a writer on a given field.

*definition 6*
Let **a** be a bag of access vectors, and I an instance of A, then we define the *read* and *write control vector* (RCV and WCV) of I as N-uples of integers such that:
$\forall i: 1 \leq i \leq N,$

$$rcv_i = \sum_{DAV_{A,OP} \in \mathbf{a}} \begin{cases} 1 \text{ if } m_i = \text{Read} \\ 0 \text{ otherwise} \end{cases} \quad wcv_i = \sum_{DAV_{A,OP} \in \mathbf{a}} \begin{cases} 1 \text{ if } m_i = \text{Write} \\ 0 \text{ otherwise} \end{cases}$$

The following lemma states that a bag of commutative operations can be summarized by the read and write control vectors. It is also the invariant of the forthcoming monitor.

*lemma 1*
**a** is pairwise commutative if and only if:
$\forall i: 1 \leq i \leq N, (rcv_i \neq 0 \Rightarrow wcv_i = 0) \text{ and } (wcv_i \neq 0 \Leftrightarrow wcv_i = 1)$

Finally, we introduce our "locking" mechanism: a monitor, and also extend it in order to eventually offer conditional commutativity by means of downgrading. In addition, this extension fits nicely into a general and previously introduced framework [10].

Access vectors are fairly conservative since they describe the most restrictive pattern that the execution of an operation could ever reach. In particular, access vectors can represent the sum of exclusive paths. This limitation can be overcome by downgrading, at the expense of, first, generating a dynamic access vector when an operation is executed, then, executing a second control. Downgrading from access vectors to dynamic access vectors (slightly) increases parallelism, and saves further space in the log.

```
monitor ATupleProtocol is
var
  ReadVector:  array [1..N] of natural;
  WriteVector: array [1..N] of boolean;
  Blocked:     array [1..N] of queue of
               tuple
                  Op: op;
                  IsFieldReader: boolean;
               end tuple;
procedure UnblockAny;
  if not Blocked[i].empty
  then if Blocked[i].first.IsFieldReader
          then ReadVector[i] := 1;
          else WriteVector[i] := true;
          end if;
          Blocked[i].dequeue(NextOp);
          signal(NextOp);
  end if;
procedure UnblockReader;
  if  not Blocked[i].empty and then
       Blocked[i].first.IsFieldReader
  then ReadVector[i] += 1;
       Blocked[i].dequeue(NextOp);
       signal(NextOp);
  end if;
entry-point InControl (in NewOp: op) is
  for i in 1..N do
    NewOp.DynamicDAV[i] := Null;
    if NewOp.DAV[i] = Read
    then if WriteVector[i]
            then Blocked[i].enqueue(NewOp,true);
                 wait;
                 UnblockReader;
            else ReadVector[i] += 1;
            end if;
    elsif NewOp.DAV[i] = Write
    then if WriteVector[i] or (ReadVector[i] ≠ 0)
            then Blocked[i].enqueue(NewOp,false);
                 wait;
            else WriteVector[i] := true;
            end if;
    end if;
  end for;

entry-point OutControl (in NewOp: op) is
  for i in reverse 1..N do
    if(NewOp.DAV[i] = Read) and
      (NewOp.DynamicDAV[i] = Null)
    then ReadVector[i] —= 1;
         if ReadVector[i] = 0
         then UnblockAny;
         end if;
    elsif (NewOp.DAV[i] = Write) and
          (NewOp.DynamicDAV[i] ≠ Write)
    then WriteVector[i] := false;
         if NewOp.DynamicDAV[i] = Read
         then ReadVector[i] := 1;
              UnblockReader;
         else UnblockAny;
         end if;
    end if;
  end for;
entry-point CommitOrReject (in NewOp: op) is
// Called either after execution of the inverse of
NewOp and then OutControl if it is a reject, or after
a commit
  for i in reverse 1..N do
    if NewOp.DynamicDAV[i] = Read
    then ReadVector[i] —= 1;
         if ReadVector[i] = 0
         then UnblockAny;
         end if;
    elsif NewOp.DynamicDAV[i] = Write
    then WriteVector[i] := false;
         UnblockAny;
    end if;
  end for;
init
  for i in 1..N do
    ReadVector[i] := 0;
    WriteVector[i] := false;
    Blocked[i] := Ø;
  end for;
end monitor.
```

Figure 1. A monitor for controlling strong-commutative accesses to tuple-based ADTs

The execution of an operation is divided into three steps: Before executing the new operation, an InControl, (the first entry-point in Figure 1), is executed. It utilizes the access vector to check commutativity of the incoming operation with previously executed or in execution ones.

Next, and if the operation is not blocked, it is executed in full parallelism with other commutative operations, i. e., execution is done outside the monitor. While it is in execution, a dynamic direct access vector (DynamicDAV) is constructed: whenever a field is actually modified, the corresponding access mode is set to Write; when a field is used in an expression, its access mode is either set to Read, or maintained to Write. These assignments are inserted in the object code at compile-time.

Finally, a second control, the OutControl, is responsible for restarting transactions which have been suspended on *a priori* possibilities of conflicts, i. e., which do not commute with respect to their access vectors but do with regard to their dynamic access vectors.

A third entry point is given to unblock operations after a reject or a commit. We rely on strict two-phase locking [7] at the transaction level.

The rationale for this extension is to offer *conditional* commutativity. In effect, with dynamic access vectors, two operations may commute for some particular values of the fields and/or the in-parameters; the monitor provides a special form of conditional commutativity, where conditions are *not* given by the programmer but discovered again and again by the monitor. (Note that if downgrading is not demanded, access vectors can be translated into mere access modes [9], which eliminates time and space overheads.)

Recovery uses dynamic access vectors to log the fields which have been actually modified.

The following two lemmas, one for downgrading and another for recovery, follow from the observation that $DAV_{A,OP\text{-}1} \leq DynamicDAV_{A,OP} \leq DAV_{A,OP}$ in the sense that each component of an access vector is less than or equal to its counterpart in the following direct access vector.

*lemma 2*
Let **a** be a bag of pairwise commutative operations on A, and OP any operation in **a**, then **a'**, obtained by substituting $DynamicDAV_{A,OP}$ for $DAV_{A,OP}$, is also pairwise commutative.

Lemma 2 states that downgrading from access vectors to dynamic access vectors does not invalidate serializability.

*lemma 3*
Let **a** be a bag of pairwise commutative operations on A, and OP any operation in **a**, then **a'**, obtained by substituting $DAV_{A,OP\text{-}1}$ for $DynamicDAV_{A,OP}$, is also pairwise commutative.

Lemma 3 states that inverse operations need not to be controlled: The third entry-point is called solely to unblock operations.

Therefore, the monitor of figure 1 is correct. It is also fair and works in constant time.
Correctness is obtained by the invariance of the property given in lemma 1.
Fairness does not imply strict FIFO policy since there are several queues to cross but not all.
Constant time is obtained by a classical parallel programming technique which consists in letting each operation in a queue be responsible for restarting the following, (rather than restarting a whole group of readers.) Constant time is also achievable by compatibility even if there are several access modes, e. g., N, IS, IX, S, SIX, and X of [8]. In that case, the control is achievable in O(m) where m is the number of access modes. Note, however, that the set of access modes must form a lattice, and that there can be up to $2^m$ combinations of access modes if each basic mode is incomparable with each other [9].

This version of the monitor assumes that there is at most one operation per transaction! The implemented version takes into account multiple operations acting on behalf of a common transaction. This would have complicated unnecessarily the presentation.

More importantly, there is one case for which downgrading is of no use. Let us imagine that there is a Write/Write dependency between two operations and that the one which is executed downgrades to Read, then the other remains blocked, though it *might* also downgrade. Introducing a restricted form of *relative recoverability* [3] seems to be a good solution. In that way, the second operation is authorized to execute; if it downgrades then they actually commute, otherwise a reject dependency is established. Even if they do not commute, both operations can be rejected independently; however, commits must take place in the dependency order.

## 4. COMPARISON WITH PREVIOUS WORKS

In the literature, access vectors are foundable as soon as in [7]: They were proposed in conjunction with predicative locking. Both predicative locking and access vectors were abandoned in System R. The former was withdrawn for performance reasons. The latter may have been eliminated due to the overhead of generating, at run-time, the access vectors of any request, and the overhead of locking with access vectors of varying length. First, we note that access vectors are entirely determined at compile-time with ADTs since every operation is known. Next, the first normal form requirement in relational databases achieves a rough form of access vectors. Effectively, complex objects spread over several relations which are, from the concurrency control point of view, separate entities, that is, locked separately. Then, the main practical reason for using access vectors is that *representing a complex object as an ADT and locking it as a whole will achieve less parallelism than in relational databases*! Even if that very important reason has not been pointed out in the literature, several other proposals introduce access vectors.

First, [13] treats a kind of strong commutativity from the recovery point of view. Modified fields are detected exclusively at run-time. The reader may have a look at a tree a nineteen propositions for recovering from rejects or crashes.

Next, [2] proposes a similar technique which is also divided into a static analysis and a dynamic one. However, the static phase requires semantic knowledge which is much harder to automate than our syntactic analysis. In the dynamic phase, the complex structure of the fields is taken into consideration, whereas we concentrate on root fields. Then, intersections of "affected-sets" are more expensive than comparisons of access vectors.

Then, very recently [1] introduces also access vectors in object-oriented databases. When invoking a method, first, the method is locked not to be modified or deleted concurrently, next, each field used by the method is locked individually, then, the method is executed. In spite of its similarity, this scheme was proposed with concurrent updates of the definitions of classes in mind: neither downgrading, nor recovery are investigated. Besides, object-oriented databases places other constraints which have to be adequately studied.

Lastly, we can parallel our technique with the methods of predeclared supersets, often used for preventing deadlocks and allowing non-two-phase locking, (e. g., [5, 6]). It is in general not feasible to determine *a priori* the sets of read- and write-accessed items in the whole database, whereas we concentrate on encapsulated data for which any access is done and controlled through known operations.

## 5. CONCLUSION AND ISSUES

Based on theoretical results [11], but also justified by practical considerations (in section 6), we proposed a simple concurrency control and recovery technique to achieve automated conditional commutativity on tuple-based ADTs. What makes this technique so attractive is that it does not require sophisticated analysis of the operations. On the one hand, our method stands between the old compatibility criterion and conditional commutativity, and on the other hand, it offers fine-granularity before- and/or after-image logging. The execution of an operation can be divided into a control phase which requires atomicity and the actual execution of the operation which can be done in *full parallelism* with other commutative operations or inverse operations of rejected transactions.

One of the major fields of application is certainly object-oriented databases: first, classes are closely related to ADTs; secondly, databases are multi-user environments; thirdly, classes and methods are expected to be frequently added, removed or updated which means that *ad hoc* commutativity relations cannot be given repeatedly. From the recovery point of view, another open issue is that neither multi-level transactions [18, 19], nor ARIES [12] are adapted because neither let two modifying operations run really concurrently on the same page, (locked temporarily in exclusive mode.) Also, access vectors can serve other purposes than concurrency control and recovery, e. g., constraints and authorization. At last, an interesting issue is to determine when a representation is optimum, i. e., allows maximal concurrency and minimal recovery with respect to a set of operations. (It is obvious that using non redundant fields is a necessary condition for recovery, hence, normalization techniques have to be employed.)